\documentclass[12pt, letterpaper]{article}

\usepackage{amsmath,amssymb}
\usepackage{cite}
\usepackage{fancyhdr}
\usepackage[top=1in, bottom=1in, left=1in, right=1in]{geometry}
\usepackage{graphicx}
\usepackage{hyperref}

\numberwithin{equation}{section}
\date{}

\title{{\rm\footnotesize \qquad \qquad \qquad \qquad \qquad \ \qquad \qquad \qquad \ \ \ \ \ \                  UTTG-31-14\ TCC-032-14     RUNHETC-2013-17     
SCIPP 13/11}\vskip.5in     Holographic Inflation and The Low Entropy of the Early Universe}
\author{Tom Banks\\
NHETC and Department of Physics \\
Rutgers University, Piscataway, NJ 08854-8019\\
{\it and}\\
Department of Physics and SCIPP\\
University of California, Santa Cruz, CA 95064\\
E-mail: \href{mailto:banks@physics.rutgers.edu}{banks@physics.rutgers.edu}}
\begin{document}

\maketitle
\thispagestyle{fancy} 
\begin{abstract}
This is a completely rewritten version of the talk I gave at the Philosophy of Cosmology conference in Tenerife, September 2014, which incorporates elements of my IFT Madrid Anthropics Conference talk.  The original was too technical.  The current version uses intuitive notions from black hole physics to explain the model of inflationary cosmology based on the Holographic Space Time formalism.  The reason that the initial state of the universe had low entropy is that more generic states have no localized excitations, since in HST, localized excitations are defined by constraints on the fundamental variables.
The only way to obtain a radiation dominated era, is for each time-like geodesic to see an almost uniform gas of small black holes as its horizon expands, such that the holes evaporate into radiation before they collide and coalesce.  Comparing the time slicing that follows causal diamonds along a trajectory, with the global FRW slicing, one sees that systems outside the horizon had to undergo inflation, with a number of e-folds fixed by the present and inflationary cosmological constants, and the black hole number density on FRW slices just after inflation ends.  These parameters also determine the size of scalar and tensor metric perturbations and the reheat temperature of the universe.  I sketch a class of explicit finite quantum mechanical models of cosmology, which have these properties.  Physicists interested in the details of those models should consult a recent paper\cite{holoinflation3}. 
\normalsize \noindent  \end{abstract}


\newpage
\tableofcontents
\vspace{1cm}

\section{Introduction}

The formalism of Holographic Space-time (HST) is an attempt to write down a theory of quantum gravity which can treat space-times more general than those accessible to traditional string theory.  String theory, roughly speaking treats space-times which are asymptotically flat or Anti-deSitter.  As classical space-times, these contain infinite area causal diamonds on which strict boundary conditions are imposed.  The corresponding quantum theory has a unique ground state, and the existing formalism describes the evolution of small fluctuations around that ground state in terms of evolution operators involving the infinite set of possible small fluctuations at the boundary.  Local physics is obscure in the fundamental formulation of the theory.   It emerges only by matching the fundamental amplitudes to those of an effective quantum field theory, in a restricted kinematic regime.  In the AdS case, one must also work in a regime where the AdS radius is much larger than the length scale defined by the string tension.  That string length scale is bounded below by the Planck length.  In regimes where the two scales are close, there are no elementary stringy excitations.  

The HST formalism works directly with local quantities.  Its important properties are summarized as follows.

\begin{itemize}

\item The fundamental geometrical object, a time-like trajectory in space-time, is described by a quantum system with a time dependent Hamiltonian. $4$ times the logarithm of the dimension ( $=$ entropy) of the Hilbert space of the system is viewed as the quantum avatar of the area of the holographic screen of the maximal causal diamond along the trajectory. The causal diamond associated with a segment of a time-like trajectory is the intersection of the interior of the backward light cone of the future endpoint of the segment, with that of the future light cone of the past point.  The holographic screen is the maximal area surface on the boundary of the diamond. 

When the entropy of Hilbert spaces is large, space-time geometry is emergent.  The case of infinite dimension must be treated by taking a careful limit.  The Hilbert space comes with a built in nested tensor factorization:  ${\cal H} = {\cal H}_{in} (t) \otimes {\cal H}_{out} (t)$.
$t$ is a discrete parameter, which labels the length of a proper time interval along the trajectory.  ${\cal H}_{in} (t)$ is the Hilbert space describing the causal diamond of that interval. ${\cal H}_{out} (t) $ describes all operators, which commute with those in the causal diamond.  We adopt the prescription from QFT that space-like separation is encoded in commutivity of the operator algebra of a causal diamond, with that associated to the region of space-time that is space-like separated from that diamond.  The holographic bound on the entropy of a finite area diamond allows us to state this in terms of simple tensor factorization.  The Hamiltonian of the system {\it must } be time dependent, in order to couple together only DOF in ${\cal H}_{in} (t)$ for time intervals shorter than $t$.  

\item Time is fundamental but relative in HST, while space-time is emergent.
By relativity of time, we mean that each time-like line has its own quantum description of the world.  Space-time is knit together from the causal diamonds of all intervals along a sufficiently rich sampling of trajectories.  For each pair of diamonds, Lorentzian geometry gives a maximal area diamond in the intersection.  The quantum version of this notion is the identification of a common tensor factor in the Hilbert spaces of these two quantum systems. The initial conditions and Hamiltonians of the two systems must be such that the density matrices prescribed by the two systems for that common tensor factor, are unitarity equivalent.  This is an infinite number of constraints on the dynamics, and the requirement is quite restrictive.  For example, in asymptotically flat space-time, the requirement of Lorentz invariance of the scattering matrix is a consequence of these consistency conditions. Note that in HST geometry is an emergent property of quantum systems, but the metric is not a fluctuating quantum variable.  The causal structure and conformal factor (which determine the metric) are determined by the area law and the overlap rules, which are not operators in the Hilbert space.  

\item The fact that geometry is not a quantum variable fits very nicely with Jacobson's derivation of Einstein's equations\cite{ted} as the hydrodynamics of a quantum system whose equation of state ties entropy to geometry via the area law for causal diamonds.  Hydrodynamic equations are classical equations valid in high entropy quantum states of systems whose fundamental variables have nothing to do with the hydrodynamic variables (the latter are ensemble expectation values of quantum operators).  There is only one situation in which quantized hydrodynamics makes sense:  small, low energy fluctuations around the ground state (of a system that has a ground state).  This accounts for the success of quantum field theory in reproducing certain limiting boundary correlation functions in asymptotically flat and AdS space-times. A notable feature of Jacobson's derivation of Einstein's equations from hydrodynamics is that it does ${\it not}$ get the cosmological constant (c.c.) term.  Fischler and I have long argued that the c.c. is an asymptotic boundary condition, relating the asymptotics of proper time and area in a causal diamond.  In the quantum theory it is a regulator for the number of states.  If the c.c. is positive, the number of states is finite.  If the c.c. is negative, it determines the asymptotic growth of the density of states at high energy.  High energy states are all black holes of large Schwarzschild radius.  In HST, the value of the c.c. is one of the characteristics that determines different {\it models} of quantum gravity, with very different Hamiltonians and fundamental DOF\footnote{See the discussion of meta-cosmology below for a model that incorporates many different values of the c.c. into one quantum system.}.

\item In four non-compact dimensions, the variables of quantum gravity are spinor functions $\psi_i^A (p)$ ($p$ is a discrete finite label, which enumerates a cutoff set of eigenfunctions of the Dirac operator on compact extra dimensions of space).  These label the states, described by local flows of asymptotically conserved quantum numbers, through the conformal boundary of Minkowski space time.  The Holographic/Covariant Entropy Principle, is implemented on a causal diamond with finite area holographic screen by cutting off the Dirac eigenvalue/angular momentum on the two sphere.  The fundamental relation is $$\pi (RM_P)^2 = L N(N + 1) , $$ where $R$ is the radius of the screen, $L$ the number of values of $p$ and $N$ the angular momentum cut-off. $i$ and $A$ range from $1$ to $N$ and $N + 1$ respectively.

\item We have constructed\cite{holomink} a class of models describing scattering theory in Minkowski space.  The basic idea of those models is that localized objects are described by constrained states, on which of order $EN$, with $E \ll N \rightarrow\infty$, matrix elements of the square matrices $M_{i}^j (p,q) \equiv \psi^{\dagger\ j}_A (p) \psi_i^A (q) $ vanish, as the size of the diamond goes to infinity.  The Hamiltonian has the form \begin{equation} H_{in} (N) = E + \frac{1}{N^2} {\rm Tr}\ P(M) , \end{equation} where $P$ is a polynomial of $N$ independent order $\geq 7$.  The quantity $E$ defining the constraint is an asymptotically conserved quantum number. The constraints imply the the matrices can be block diagonalized and, when combined with the large $N$ scaling of the Hamiltonian, the blocks are free objects: the Hamiltonian is a sum of commuting terms, describing individual blocks.  It can be shown that in the limit in which the individual blocks have large size, these objects are supersymmetric particles.  For generic choice of $P(M)$ the long range interactions have Newtonian scaling with energy and impact parameter.  All of these models have meta-stable excitations with the properties of black holes, which are produced in particle scattering and decay into particles. Scattering amplitudes in which black holes are not produced can be described by Feynman like diagrams, with vertices localized on the Planck scale.  We have not yet succeeded in imposing the HST consistency conditions for trajectories in relative motion, which we believe will put strong constraints on $P$ and on the spectrum of allowed particles\footnote{The spectrum is encoded in the commutation relations of the variables $\psi_i^A (p)$.}.

\item It's easy to convert the models above into models of de Sitter space, by simply keeping $N$ finite.   This automatically explains both the de Sitter entropy and temperature, the latter because the definition of a particle state of energy $E$ involves constraining $EN$ of the $o(N^2)$ DOF on the horizon.
The probability of having such a state, within the random ensemble we call the dS vacuum state\footnote{It's been known since the seminal work of Gibbons and Hawking\cite{gibbhawk}, that the dS ``vacuum state" is actually a high entropy density matrix.} is $e^{- EN} $. This says dS space has a temperature proportional to the inverse of its Hubble radius.  The fact that localized energy corresponds to a constraint on the degrees of freedom in dS space is already evident from the form of the Schwarzschild-de Sitter black hole solution.  It has two horizons, given by the roots of $( 1 - 2GM/r - r^2/R^2)$.  It's easy to verify that the sum of the areas of the two horizons is minimized at $M = 0$, and when $M$ is small, the entropy deficit is that expected from a Boltzmann factor $e^{ - 2\pi R M}$.  

\item We have constructed\cite{holocosmath}\cite{holoinflation123} a fully consistent quantum model of cosmology, in which the universe is described by a flat Friedmann Robertson Walker (FRW) metric with a stress tensor that is the sum of a term with equation of state $p = \rho$ and a term with $p = - \rho$. The geometry, as anticipated by Jacobson, is a coarse grained thermodynamic description, valid in the large entropy limit, of the quantum mechanics of the system.  Homogeneity, isotropy and flatness are realized for arbitrary initial states.  Homogeneity and isotropy are the only obvious ways to satisfy the consistency conditions between the descriptions of physics along different trajectories, when each trajectory is experiencing randomizing dynamics.  Flatness follows from an assumption of asymptotic scale invariance for causal diamonds much larger than the Planck scale but much smaller than the Hubble scale of the c.c. .  The c.c. itself is an input, basically a declaration that we stop the growth of the Hilbert space, but allow time evolution to proceed forever, with a fixed Hamiltonian, which has entered the scaling regime describing the asymptotics of the $\Lambda = 0$ model.
Note by the way that the initial singularity does not appear.  The geometric description is valid only in the limit of large entropy/large causal diamonds and the singularity is an extrapolation of this limiting behavior back to a time when the causal diamond is Planck size.  We have called this model Everlasting Holographic Inflation (EHI).  It has an infinite number of copies of a space-time which is asymptotically a single horizon volume of dS. Unlike field theoretic models of eternal inflation, the different horizon volumes are constrained to have identical initial conditions and may be viewed as gauge copies of each other. In HST stable dS space is a quantum system with a finite dimensional Hilbert space\cite{tbwf} . This model has no local excitations, except those which arise as thermal fluctuations in the infinite dS era. It is important to note that the EHI model and the more realistic model described below are not the same as the HST model of stable dS space.  The latter model has infinite, rather than semi-infinite proper time intervals, and a Hamiltonian, for each trajectory, which satisfies $H(T) = H(-T)$ for each proper time.  In the limit of infinite dS radius, it approaches the HST model for Minkowski space.  Both the EHI model and our semi-realistic Holographic Inflation model use the same time dependent Hamiltonian $H(T)$, where $T$ runs over a semi-infinite interval.  There is no time reflection symmetry in the system.   EHI and HI differ only in their boundary conditions, with the latter having fine tuned boundary conditions, which guarantee an era where localized excitations are approximately decoupled from horizon DOF.  In EHI, {\it despite its intrinsic time asymmetry}, the universe is always in a generic state of its Hilbert space at all times, and local excitations, which are of low entropy, because they are defined by constrained states on which large numbers of $\psi_i^A (p)$ vanish, arise only as sporadic thermal fluctuations.

\item {\it Perhaps the most important difference between HST and QFT lies in the counting of entropy.  In HST the generic state of the variables in a causal diamond of holoscreen area $ \sim R^2$, has no localized excitations: all of the action takes place on the horizon, with a Hamiltonian that is not local on the holographic screen.  Bulk localized states are constrained states in which of order $ER$ of the $o(R^2)$ variables are set equal to zero.  $R$ is the holoscreen radius, and $E \ll R$ (all these equations are in Planck units) is the approximately conserved energy (it becomes conserved in the limit $R$ goes to infinity).  It is only for these constrained states that QFT gives a good description of some transition amplitudes.}

\end{itemize}

Our model of the universe we live in, the topic of this talk, proceeds from a starting point identical to that of the model discussed in the penultimate bullet point.  However, we consider a much larger Hilbert space, for a single trajectory, which includes many copies of a single inflationary horizon volume.  This corresponds to the growth of the apparent horizon, after inflation, in conventional inflationary models.
This model contains elements of the conventional narrative about cosmological inflation, so we call it the Holographic Inflation model.  The purpose of the inflationary era in this model is quite different from that which inflation serves  in field theory models.  Homogeneity, isotropy and flatness are natural in HST.  We need fine tuning of initial conditions in order to get local excitations, and this is the purpose that the inflationary era serves.  I'll comment below on the degree of fine tuning required compared to field theoretic inflation, but the most important point is that our fine tuning is the minimal amount required to get localized excitations, so the very crudest kind of anthropic reasoning, a {\it topik\`{e}sthrophic} restriction (from the Greek word for locality), says that, within the HST formalism, a universe with this amount of fine tuning of initial conditions is the only kind that can ever be observed. The only assumption that goes into this is that any kind of observer will require the approximate validity of local bulk physics.  It does not require the existence of human beings, or anything like conventional biology.  

The derivation of a macroscopic world from quantum mechanics, a world in which the ordinary rules of logic and the notion of decoherent histories are valid, relies on the existence of macroscopic objects with macroscopic moving parts.
In the HST model of quantum cosmology, a typical macroscopic object is the entire apparent horizon, slightly less typical ones are localized black holes.  None of these have complex webs of semi-classical collective coordinates.  We know how to derive the existence of complex macro-objects in quantum field theory, even with an ultra-violet cutoff, but in HST quantum states approximately describable by QFT are highly atypical.  If we want a universe in which such states appear as anything but ephemeral thermal fluctuations, we {\it must} impose constraints on the initial conditions.  Thus, in HST, the reason the universe began in a low entropy state, is that this is the only way in which the model produces a complex, approximately classical world.

\section{The Holographic Inflation Model}

Our cosmology begins on a space-like hypersurface, called the Big Bang, which has the topology of three dimensional flat space. A sampling of time-like trajectories in the emergent space-time, are labelled by a regular lattice on the space-time, whose geometry is irrelevant.  Probably a general three dimensional simplicial complex, with the simplicial homology of flat space is sufficient.  The quantum avatar of each trajectory is an independent quantum system, whose Hilbert space is finite dimensional if the ultimate value of the c.c. is positive.

We incorporate causality into our quantum system, by insisting that the Hamiltonian is proper time {\it dependent} along the trajectory and has the form
\begin{equation} H( t_n) = H_{in} (t_n) + H_{out} (t_n) . \end{equation}  At proper time $t_n$ the horizon has area 
\begin{equation} \pi (R_n M_p)^2 = L n ( n + 1) . \end{equation}
$H_{in} (t_n) $ is constructed from the matrices $M$ built out of the subalgebra of $\psi_i^A $ with indices $i \leq n$ and $ A \leq n + 1$ .  $H_{out} (t_n)$ depends only on the rest of the variables.  In our simplest models, we will not have to specify $H_{out}$, but it will be crucial to our Holographic Inflation Model.  

All of our models choose exactly the same sequence of Hamiltonians, and the same initial state, for each trajectory in our lattice.  The initial state is however, completely arbitrary, so there is no fine tuning of initial conditions.  Along each trajectory, the Hamiltonian $H_{in} (t_n)$ is the trace of a polynomial, $P$, in the matrices $M (t_n)$.  The coefficients in $P$ are chosen randomly at each time $t_n$. Systems like this almost certainly have fast scrambling behavior\cite{haydenstanford}. The maximal order of the polynomial is fixed and $\ll N$, but we don't yet know what it is beyond a bound ${\rm deg} P \geq 7$.   This bound is required to get the proper scaling of Newton's law in regimes that are approximately flat space\cite{holonewton}.
Finally, we require that, when $n$ is large, but $\ll N$, the Hamiltonian approach that of a $1 + 1$ dimensional C (onformal) F(ield) T(heory) on a half line, with central charge of order $n^2$.  The system is obviously finite dimensional, so this statement can only make sense in the presence of UV and IR cut-offs on the CFT .  We choose them to be $\Lambda_{UV} \sim 1/n$,  $V_{IR} \sim L n $, and also insist that $L \gg 1$ so that the CFT behavior will be manifest in the presence of the cutoffs.

The fast scrambling nature of the dynamics implies that we can make thermodynamic estimates of the expectation value of $H_{in}$ and the entropy of the time averaged density matrix $$ E \sim L n , $$ $$ S\sim L n^2 . $$  The volume of the bulk region inside the horizon at this time scales like $n^3$, so the energy and entropy densities scale like
$$ \rho \sim n^{-2} , $$
$$ \sigma \sim n^{-1} = \sqrt{\rho} .$$ If we recall that in a flat FRW metric the horizon size $n$, scales like the cosmological time $t$ we recognize the first of these equations as the Friedmann equation and the second as the relation between entropy and energy densities for an equation of state $ p = \rho$.  This geometric description was to be expected, from Jacobson's argument\cite{ted} showing that Einstein's equations are the hydrodynamics of systems obeying the area/entropy connection.
Note that the geometric/hydrodynamic description should not be extrapolated into the low entropy regime of small $n$, so that the cosmological singularity of the FRW cosmology is irrelevant.   Note also that the quantum mechanics of this system is in no sense that of the quantized Einstein equations. Quantized hydrodynamics is a valid approximation for describing small excitations of a system around its ground state.  In the HST models, the early universe is very far from its ground state, and does not even have a time independent Hamiltonian\footnote{In passing we note that the only known quantum gravitational systems with a ground state are asymptotically flat and anti-de Sitter space-times.  The proper description of these is String Theory-AdS/CFT, and certain amplitudes in the quantum theory {\it are} well approximated by QFT. IMHO, string theorists and conventional inflation theorists make a mistake in trying to extrapolate that approximation to the early universe.}

In the EHI models, the quantum systems along different trajectories are knit together into a space-time, by specifying at each time $t_n$ that the maximal causal diamond in the intersection between causal diamonds of two trajectories that are $D$ steps apart on the lattice, is the tensor factor in each Hilbert space, on which $H_{in} ( t_n - D)$ acted.  The choice of identical Hamiltonians and initial state for each trajectory insures that the density matrix on this tensor factor at time $t_n$ is independent of which trajectory we choose to view it from.  This infinite set of consistency conditions is the fundamental dynamical principle of HST.  The locus of all points $D$ steps away on the lattice is identified by this choice as a set of points on the surface of a sphere in the emergent space-time, because of the causal relations.  The fact that the dynamics is independent of the point is rotation invariance on that sphere, and this is consistent with the fact that our fundamental variables transform as a representation of $SU(2)$ and the Hamiltonian is rotation invariant.  We see that homogeneity, isotropy and flatness are properties of the space-time of this model, which are independent of the choice of initial state.  In our more realistic models of the universe, in which local objects emerge from a choice of fine tuned initial conditions, we do not yet have a solution to the the consistency conditions for trajectories with Planck scale spacing, but homogeneity and isotropy play a crucial role in satisfying a coarse grained verse of the consistency conditions for trajectories whose spatial separation is of order the inflationary horizon size.

As $n \rightarrow N$, we need to change the rules only slightly.  Proper time is now decoupled from the growth of the horizon.  We model this by allowing the system to propagate forever with the Hamiltonian $H_{in} (N)$.  In addition, we do an (approximate) conformal transformation on the Hamiltonian, rescaling the UV and IR cutoffs so that the total Hamiltonian is bounded by $1/n \rightarrow 1/N$.  This is analogous to the transformation from FRW to static observer coordinates\cite{sussetal} in an asymptotically dS universe, and is appropriate because we are postulating a time independent Hamiltonian.  Finally we have to modify the overlap rules, to be consistent with the fact that individual trajectories have a finite dimensional Hilbert space.  The new overlaps are the same as the old ones, except that points that are more than $N$ steps apart on the lattice never have any overlap at any time.  The asymptotic causal structure is then that of dS space, with Hubble radius satisfying $\pi (R_H M_P)^2 = L N (N + 1) .$

The simplest metric interpolating between the $p = \rho$ and $p = - \rho$ equations of state is $a(t) = \sinh^{1/3} (3t/R_H)  $, which is an exact solution of Einstein's equations with a fluid that has two components with these equations of state.  All of the geometric information in our model is consistent with this ansatz for the metric.   The space-time of our model contains no localized excitations.  At all times, all DOF are localized on the apparent horizon, in a completely thermalized state, obeying none of the constraints which characterize bulk localized systems in HST\cite{holomink} .  This EHI model, is not a good model of the universe we inhabit, although it is, according to the rules of HST, a perfectly good model of quantum gravity.  Localized excitations will occur only as ephemeral thermal fluctuations in the eternal dS phase of this cosmology.

\subsection{A More Realistic Description of the Universe}

The description of localized objects in HST was worked out in a series of papers\cite{holomink} devoted to scattering theory in Minkowski space-time.   Here we summarize the results and explain how they're used to construct the HST version of inflationary cosmology.  The variables $\psi_i^A (p)$ are sections of the Dirac-cutoff spinor bundle on the two sphere, so the matrices $M^j_i (p,q)$ are sections of the bundle of differential forms on the sphere.  Two forms can be integrated over the sphere and the fuzzy analog of integration is the trace, which we have used in constructing our model Hamiltonians.  Expressions involving a trace of a polynomial in $M$ are invariant under unitary conjugation and this invariance converges to the group of measure preserving transformations on the sphere (we do not require them to be smooth).  Saying that a matrix is block diagonal in some basis, can be interpreted as saying that the corresponding forms vanish outside of some localized region on the sphere.  In our quantum mechanics, the matrix elements of $M$ are operators, so a statement that some of them vanish is a {\it constraint on the states}.  If this constraint is {\it approximately} preserved under the action of the Hamiltonian taking us from one causal diamond to the next, then this defines the track of a localized object in space-time.  

This connection between localization, and constraints that put the system in a low entropy state, is supported by a piece of evidence from classical GR.  An object in dS space, localized in a region much smaller than the dS radius, will have a dS Schwarzschild field.  The local entropy will be maximized if the object is a black hole, filling the localization region, but independently of that choice,
the entropy of the horizon shrinks, so that the total entropy of the system is less than that of empty dS space.  Recall that empty dS space is a thermal system, with DOF that must be considered to live on the cosmological horizon.  The idea that a localized object of energy $E$ is a constraint on a system with $o(N^2)$ DOF, which freezes $o(EN)$ of them, explains the scaling of the temperature with dS radius.  In \cite{holomink} we showed that the same idea explains the conservation of energy in Minkowski space, as well as the critical impact parameter at which particle scattering leads to black hole production, the temperature/mass/entropy of black holes, and even the scaling of large impact parameter scattering with energy and impact parameter (Newton's Law).

All of these results are quite explicit calculations in our quantum mechanical matrix models, and they mirror scaling laws usually derived from classical GR.
It is crucial to all of them that the block diagonal constraint on matrices, gives us a finite number of blocks of size $K_i$, and one large block of size $N - K \gg 1$, where $K =\sum K_i$, and turns out to be the conserved energy.  The individual $K_i$ represent the amount of energy going out through the horizon in different angular directions.  Energy is only conserved in the limit $N \rightarrow \infty$, since in that limit the Hamiltonian cannot remove $O(KN)$ constraints.   A final result from the matrix model shows that large impact parameter scattering amplitudes scale with energy and impact parameter as one would expect from Newton's law.   Again, the existence of a vast set of very low ($o(1/N)$) energy degrees of freedom (DOF), which do not have a particle description, is crucial to this result.  In passing, I remark that these DOF resolve the firewall paradox of AMPS\cite{ampsfw3}.

The correlation between localization and low entropy is the key to answering Penrose's question\footnote{Which of course goes back to Boltzmann.  Penrose was the first to raise the issue in the context of the General Relativistic theory of gravity.} of why the initial state of the universe had low entropy.  We've already seen that an unconstrained initial state of the universe leads to the EHI cosmology, whose only localized excitations are low entropy thermal fluctuations late in the dS phase of the universe.  

The maximal entropy state with localized excitations, is one in which those excitations are black holes.  As the horizon expands one may encounter more black holes or the process may stop at some fixed horizon size.  The maximal set of black holes for a fixed horizon size $R$ in Planck units is constrained by the inequality
\begin{equation} \sum K_i \leq R .\end{equation}  Note that this constraint can be derived {\it either} from the geometric requirement that the Schwarzschild radius of the total black hole mass be smaller than the horizon size, or from the matrix model constraint that the black hole blocks of the matrix fit inside the full matrix available at that value of the horizon size.  

At this point, we must recognize the distinction between time in the matrix model, which represents the time along a particular trajectory, and FRW time.   The time slices in the matrix model always correspond to ``hyperboloids" lying between the boundaries of two successive causal diamonds, while FRW time corresponds to horizontal lines.  If we say that at some fixed causal diamond size, smaller than the dS radius of the ultimate c.c., the process of new black holes coming into the horizon stops, then trajectories far enough from the one we have been discussing will see the region around our preferred trajectory, as a collection of black holes localized in a given region.  The system will not be, even approximately, homogeneous on the would be FRW slices.  Some of the black holes may decay into radiation, but many will be gravitationally bound, and coalesce into large black holes, which have lower probability of decay.  The universe will not look anything like our own, and it is unlikely to ever produce complex structures, which could play the role of ``observers".  It is often said that once a single galaxy forms, the question of the evolution of observers is independent of the rest of the cosmos, but that statement depends on defining a galaxy as a gravitationally bound structure whose constituents are primarily composed of baryonic matter.  In these HST models of the universe, the primordial constituents of the universe are black holes.  Matter, baryonic or otherwise, must be produced in black hole decay.

Although we have not explored all such inhomogeneous scenarios, it seems clear that a model with a fairly homogeneous black hole gas is much more likely to produce an observer ready cosmology than an inhomogeneous one.  If the gas is sufficiently dilute, and the black holes sufficiently small compared to the size of the ultimate cosmological horizon, they will all decay before they can coalesce into larger black holes. That decay is the hot Big Bang in the HST models. 

It is also, as we will see, much easier to solve the HST constraints relating the descriptions along different time-like trajectories when the universe is homogeneous and isotropic in a coarse grained way.  Exact homogeneity and isotropy are incompatible with quantum mechanics.  Our black holes are really quantum systems with finite dimensional Hilbert spaces, with a quantum state that is varying by order one on a time scale $nL_P$, the Schwarzschild radius.  Occasionally the Hamiltonian will put us in a state where the value of $n$ is effectively smaller, because some of the matrix elements vanish.  Thus, the black hole radius is to be thought of as an expectation value of an operator which is the trace of some polynomial in the matrices $M$.  It will have statistical fluctuations in the time averaged density matrix of the system.  By the usual rules of statistical mechanics these will be small and approximately Gaussian, with size 
$$ \frac{\delta n}{n} \sim \frac{1}{n} .  $$  Similarly, although the expectation value of the black hole angular momentum is obviously zero, it will have Gaussian fluctuations of the same order of magnitude.   Since these are fluctuations in collective coordinates of a large chaotic quantum system, and time averaged fluctuations at that, it is obvious that quantum interference effects in the statistics of these variables are negligible, of order $e^{- n^2}$.  

Fluctuations of the mass and angular momentum of a black hole, are fluctuations of the spin zero and spin two parts of the Weyl Curvature tensor, and thus have the properties of scalar and tensor fluctuations in cosmological perturbation theory.  We will postpone an analysis of the phenomenological implications of this remark until we have completed the sketch of our model.

Now let's return to the matrix model description of our model.  For some period of time, which we call the era of inflation, the horizon size remains fixed at $n$ and the system evolves with a time independent Hamiltonian
\begin{equation} H_{in} (n) = \frac{1}{n^2} P(M_n), \end{equation} which is the asymptotic Hamiltonian of our HEI model.  Then, the horizon begins to expand, eventually reaching the cosmological horizon $N$.   At time $t_k$, when the horizon size is $k$, some number of black holes will have come into the horizon.  If the average black hole size is $n$, the $k \times k$ matrices $M_k$, which appear in the Hamiltonian $H_{in} (t_k)$ must be in a block diagonal state, with some number of blocks $p_k \leq k/n$ of size $n$.  The black hole energy density is thus
\begin{equation} \rho_{BH} = \frac{p_k n}{k^3} \leq \frac{1}{k^2} .  \end{equation} If we choose $t_k \propto k$, as we expect for any flat FRW model, then the RHS is just the Friedmann equation for the energy density if we choose $p_k = n_{BH} \frac{k}{n} , $ with $n_{BH}$ interpreted as an initial black hole number density.  

I do not have space here to sketch the full matrix model treatment of this system, but will instead rely on the reader's familiarity with semi-classical black hole dynamics, together with the evidence from \cite{holomink} that thermalized block diagonal matrices have qualitative behavior very similar to that of black holes.   The initial number density of black holes on FRW slices, must be $ < 1/n^3 $, so that the black holes are further apart than their Schwarzschild radii.
What happens next is a competition between two processes: the growth of fluctuations in a universe dominated by an almost homogeneous gas of black holes, and the decay of the black holes.   As we will recall below, the size of the primordial scalar (mass) fluctuations in the black hole energy density is $Q = \frac{\delta\rho}{\rho} \sim\frac{1}{n\epsilon}$, where $\epsilon$ is a ``slow roll parameter", currently bounded above by $\sim 0.1$ by observation. In the black hole dominated universe, these will grow to $o(1)$ when the scale factor has grown by $n\epsilon$, which occurs in FRW time $t_{FRW} \sim (n \epsilon)^{3/2}$, which is much less than the black hole evaporation time $t_{evap} \sim n^3$.  Black holes will thus begin to combine on this time scale, potentially shutting off the evaporation process and leaving us with a universe dominated by large black holes, forever.  However, since black holes are separated by distances {\it much} larger than their Schwarzschild radii by this time, the fluid approximation does not capture the full coalescence process.  We must also estimate the infall time for two widely separated black holes in a local over-density to actually collide.  It turns out that this time is longer than the decay time as long as $n_{BH} < n^{-3} $.   So our model produces a radiation dominated universe, with a reheat temperature given roughly by the redshifted 
black hole energy density at the decay time
\begin{equation} g T^4 \sim n n_{BH} \frac{g^2}{n^6} < \frac{g^2}{n^8} . \end{equation}  Here $g$ is the effective number of massless species into which the black holes can decay.
The standard model gives $g \sim 10^2$.  Supersymmetry, particularly with a low energy SUSY breaking sector added on boosts this to $g\sim 10^3$.  The relation between $n$ and primordial fluctuations gives
\begin{equation} T_{RH} < g^{1/4}  \epsilon^2 Q^2 = 10^{-9.25} \epsilon^2 < 10^{-11} . \end{equation} This estimate, which takes into account the observations of $Q$ and the observational bound on $\epsilon$ is in Planck units and corresponds to $T_{RH} < 10^7 - 10^8$ GeV.   

What does all of this have to do with inflation?  In order to answer that, we have to return to the time $t_k$, and ask what the Hamiltonian $H_{out} (t_k)$ must be doing in order to be consistent with the fact that a new system, with $o(n^2)$ DOF is about to be added to $H_{in} (t_k) $, and also with the overlap conditions of HST.  The first of these constraints says that, at the time it comes into ``our" trajectory's horizon this system must have been completely decoupled from the rest of the DOF in the universe.   This is consistent with dynamics along the trajectory that has contained those DOF in its causal diamond since the horizon size grew to $n$, {\it if that trajectory is still experiencing inflation}.  Here again we must recall the differences between the time slicings of individual trajectories, and the FRW slicing.  The future tip of a given trajectory's causal diamond lies on a particular FRW slice, but its intersection with a spatially remote trajectory is at a correspondingly remote FRW time in the past.  In an asymptotically dS universe, where the scale factor $a(\eta )$ has a pole at some conformal time $\eta_0$, the last bit of information that comes into the horizon of a given trajectory is on an FRW slice with conformal time $\eta_0 / 2$.  Thus, for an approximately homogeneous and isotropic collection of black holes, the universe had to undergo inflation up to this conformal time.  Along each trajectory, until the time $t*$ when the future tip of its diamond hits the FRW surface $\eta_0 /2$, the Hamiltonian $H_{in} (t < t*)$ is the Hamiltonian of the EHI universe. 

The shortest wavelength fluctuations that occur in this model will have wavelength of order $n a_{NOW}/a_I $, where $a_I$ is the scale factor at the end of inflation.  These are fluctuations that came into the horizon just after inflation ends.  The largest wavelength fluctuations are those which cover the entire sky and have wavelength of order $N$.  In a conventional inflation theory we would write $e^{N_e} \geq \frac{N a_I}{n a_{NOW}} , $ whereas in the HST model this is a strict equality.  There are only as many e-folds as we can see.
 A crude estimate, given the cosmological history we have sketched, gives $N_e \sim 80$.  This is larger than the conventional lower bound $N_e > 60$, because our cosmology has a long period in which the universe is dominated by a dilute black hole gas.  Reheating does not occur immediately after inflation.
 
 \subsection{$SO(1,4)$ Invariance }
Work of Maldacena and others\cite{maldaetal} has shown that current data on the CMB can be explained in a very simple framework, with no assumptions about particular models.  Indeed, it was shown in \cite{holoinflation2} that the even the assumption that fluctuations originate from quantized fields is unnecessary.
All one needs is the approximate $SO(1,4)$ invariance we will demonstrate below.

If we consider general perturbations of an FRW metric, then as long as the vorticity of the perturbed fluid vanishes (which is an automatic consequence of the symmetry assumptions below), we can go to co-moving gauge, where the perturbations are in the scalar and spin two components of the Weyl tensor.
The conventional gauge invariant scalar perturbation $\zeta$ is equal, in this gauge, to the local proper time difference between comoving time slices, in units of the background Hubble scale
\begin{equation} \zeta = H \delta\tau = \frac{H^2}{\dot{H}} \frac{\delta H}{H} \equiv  \frac{\delta H}{\epsilon H} . \end{equation}
This equation alone, if $\epsilon $ is small, explains why we have so far seen scalar, but not tensor curvature fluctuations.  The gauge invariant measure of scalar fluctuations is suppressed if the background FRW goes through a period in which the variation of the Hubble parameter is smaller than the Hubble scale, which is the essential definition of an inflationary period.  The rest of the data on scalar fluctuations essentially corresponds to fitting the background $H(t)$.  
The HST and field theory based models have a different formula\cite{holoinflation2}\cite{bft} relating $H(t)$ to the form of the scalar two point function, but if both models have an approximate $SO(1,4)$ symmetry, they can both fit the data.  

Much is made in the inflation literature of the fact that inflationary fluctuations are approximately Gaussian, and this is supposed to be evidence that free quantum field theory is a good approximation to the underlying model.  However, Gaussian fluctuations are a good approximation to almost any large quantum system in a regime where the entropy is large.  Higher order correlation functions are suppressed by powers of $S^{- 1/2}$.  The HST model has approximately Gaussian fluctuations for these general entropic reasons, rather than the special form of its ground state.  The insistence that a huge quantum system be in its ground state, is a monumental fine tuning of initial conditions, and should count as a strike against conventional inflation models.

There is a further suppression of non-Gaussian correlations involving at least one scalar curvature fluctuation, which follows from Maldacena's squeezed limit theorem\cite{maldafluct} and approximate $SO(1,4)$ invariance .  Maldacena's theorem says that in the limit of zero momentum for the scalar curvature fluctuation, the three point correlator reduces to something proportional to the violation of scale invariance in the two point correlator.  The symmetry allows us to argue that this suppression is present for all momenta. Since all 3 point functions are suppressed relative to two point functions by a nominal factor of $10^{-5}$, and this theorem gives us an extra power of $10^{-2}$, we should not expect to see these non-Gaussian fluctuations if the world is described by any one of a large class of models, including both HST and many slow roll inflation models .

The tensor two point function is the most likely quantity to be measured in the near future.  HST and slow roll models make different predictions for the tilt of the tensor spectrum.  HST predicts exact scale invariance while slow roll inflation models have a tilt of $r/8$, where $r$ is the tensor to scalar ratio.  Since we now know that $r < 0.1$ with $95\% $ confidence, it will be difficult though not necessarily impossible to observe this difference.  Of course, if $r$
is much smaller than its observational upper bound, this observation will not be feasible.  We should however point out that HST has a second source of gravitational waves, the decay of black holes.  This will have a spatial distribution that mirrors the scalar fluctuations, and so should have the same tilt as the scalars, again disagreeing with the predictions of standard slow roll models.  It is suppressed by a factor $1/g$, the number of effective massless species into which the black holes can decay, but with no suppression for small $\epsilon$.  Finally, I'd like to point out that the PIXIE mission\cite{spergel}, will test for short wavelength primordial gravitational waves and can probably distinguish even a small tilt of the spectrum.

The tensor three point function is, in all models having approximate $SO(1,4)$ invariance, might be the largest of the non-Gaussian fluctuations, unless $r \sim .04$ or less.  It does not have the $n_S - 1$ factor from Maldacena's squeezed limit theorem.  It is by far the most interesting correlation function that humans might someday observe, since symmetry allows three distinct forms for the three point function.  Quantum field theory models predict that one of the three dominates the other two by a factor of $n > 10^6
$ while the third vanishes to all orders in inverse powers of $n$.  HST models predict two of the three form factors are of comparable magnitude, while the third appears to vanish only if a certain space-time reflection symmetry is imposed as an assumption.  Unfortunately, we will probably not be able to measure this three point function in the lifetime of any of my auditors at this conference.

We turn briefly to the derivation of approximate $SO(1,4)$ invariance of the fluctuations in the HST model.  As discussed above, the Hamiltonian acting on the Hilbert space of entropy $L n^2$ in the HEI model is the Cartan generator of an approximate $SL(2)$ algebra.  This is the statement that the early universe is described approximately by a $1 + 1$ dimensional CFT.  As DOF come into the horizon, the initial state must be constrained so that the matrices $M(kn; p,q)$, which appear in $H_{in} (kn)$  are all block diagonal, with blocks of approximate size $n$.  In order for this to be consistent $H_{out} (kn)$ must act on all of the blocks that have not yet come into the horizon, but will in the future, as a sum of independent copies of the $SL(2)$ Cartan generator, $L_0 [a]$.  This insures that these DOF will have dynamics that mirrors a horizon of fixed size.  Once the horizon size has expanded to $N = K n$, corresponding to the observed value of the c.c., all of these DOF are embedded in an $SU(2)$ covariant system, with entropy $L N^2$.  We can organize the DOF, so that $SU(2)$ invariance is preserved at all times, by choosing to define the action of $SU(2)$ so that DOF, which have come into the horizon when its size is $kn$ transform in the $[kn] \otimes [kn + 1]$ representation of $SU(2)$.   

Once the apparent horizon coincides with the cosmological horizon, we can divide the entire set of variables up into variables localized at various angles.  To visualize this, take the sphere of radius $\sim Kn = N$ and draw a spherical grid: an icosahedron with triangular faces, each of which is tiled by equilateral triangles of area $n^2$.  We call ${\bf \Omega_i}$ the solid angular coordinate of the $ith$ small triangle.  Consider the $n^2$ most localized linearly independent functions we can construct from $o(N^2)$ spherical harmonics, and make a basis which consists of these localized functions in once particular tile, and all rotations of them to different tiles.   

The total number of black holes that come into the horizon is certainly no greater than $K = N/n$, and the number which have come in when the horizon size is $kn$ is $< k $.   As we've said, the initial state wave function must be such that the matrices $M(kn)$ are block diagonal, with a number of small blocks $< k$.  We choose the wave function such that the black hole (if any) which comes in when the horizon grows from $kn$ to $(k + 1)n$ is a linear superposition of wave functions in which the black hole DOF are taken from each of the independent tiles on the cosmological horizon.  This makes a state which is rotation invariant.  We also insist that the rate at which black holes is added is uniform in time.  This rate is a parameter of the model which, in FRW slicing, is determined by the initial black hole number density, $n_{BH}$, at the end of inflation.   The fact that the objects being superposed are large quantum systems with a fast scrambler Hamiltonian and a time scale $n$, means that quantum interference terms in this superposition are negligible, of order $e^{ - n^2}$, so the prediction of the model is that there is a classical probability distribution, for finding black holes at various positions in the emergent FRW space-time.  

The homogeneous and isotropic nature of the black hole distribution, from the point of view of one trajectory, makes it easy to satisfy the HST consistency conditions between trajectories.  We simply choose the same sequence of both in and out Hamiltonians, and the same initial state, for each trajectory, and let the geometry of FRW tells us what the overlap Hilbert spaces are.  This is only a coarse grained solution of the consistency conditions because both our time steps and the spatial separations of the various trajectories are of order $n$, rather than the Planck scale.   Note that apart from the fine tuning necessary to guarantee a certain number of localized excitations the conditions of homogeneity and isotropy arise naturally, and do not require any extra tuning of the initial state.  We can certainly find other solutions of the consistency conditions with inhomogeneous distributions of black holes.  However, we've argued that these will typically lead to cosmologies in which the entire content of the universe is a few large black holes, which slowly decay back into the horizon of empty dS space.

At any rate, we can combine the local $SL(2) (a)$ groups, with the generators of rotations to construct an algebra that approximates $SO(1,4)$.  In flat coordinates, the $SO(1,4)$ algebra splits into $7$ generators whose action is geometrically obvious, and $3$ which act in a non-intuitive way.  The geometric generators are the rotations and translations of the flat coordinates, and the rescaling of the flat space coordinates combined with the translation of FRW time (rescaling of conformal time).  In terms of familiar rotations and boosts in five dimensional Minkowski space, the time translation generator is $J_{04}$, the rotations are $J_{ij}$ and the translations are $J_{+i}$, where the $+$ refers to light front time, $X^0 + X^4$ in the embedding coordinates of dS space in five dimensional Minkowski space. The action of the remaining $J_{-i}$ generators is non-linear in the flat coordinates.

In the matrix model we define $J_{ij} = \epsilon_{ijk} J_k$ to be the obvious rotation generator.  The rest of the $SO(1,4)$ generators are defined in terms of the local $SL(2) [a] $ generators.
$$J_{04} = \sum L_0 ({\bf \Omega}[a] ) , $$ and \footnote{$[a]$ is a label for a tile in the spherical grid described above.  The sums are over all tiles.}
$$J_{\pm i} = \sum L_{\pm} ({\bf \Omega} [a] ) {\bf \Omega [a]}_i . $$
These operators operate only on the tensor factor ${\cal H}_{out} (t_k) $ of the Hilbert space, describing degrees of freedom which have not yet come into the horizon.  The Hamiltonian of these DOF is a sum of non-interacting terms, as one would expect for disjoint horizon volumes in dS space. 

The operators described above satisfy the commutation relations of $SO(1,4)$ with accuracy $1/n$, when $N/n$ is large.   The density matrix of the system outside the horizon is the tensor product of density matrices $\rho (\Omega_a )$ for the individual, angularly localized, systems of $n^2$ variables. This is approximately invariant under the individual $SL(2)$ generators at fixed angle, because of the fact that we've gone through a large number of e-folds of evolution with the Hamiltonian $L_0 (a)$.  It's also invariant under permutation of the individual blocks of $n^2$ variables.  We've argued that these systems enter the horizon as an $SU(2)$ invariant distribution of black holes, if we want the model universe to have a radiation dominated era.  It follows that the distribution of black hole fluctuations on the sky of each trajectory, is approximately $SO(1,4)$ invariant.  This is what we need, to fit the data on the CMB, Large Scale Structure, and galaxy formation.

\section{Meta-Cosmology and Anthropic Arguments}

It's already obvious that our resolution of many of the problems of cosmology relies to a certain extent on what are commonly called anthropic arguments.  The resolution of the Boltzmann/Penrose conundrum of why the universe began in a low entropy state is that typical initial states, evolve under the influence of the HST Hamiltonian, into states which consist entirely of apparent horizon filling black holes, and that the system asymptotes to the density matrix of empty dS space\footnote{We will discuss more elaborate models, in which the c.c. itself is selected anthropically, below.} without ever producing localized excitations.  It is, I would claim, obvious, that any such model cannot have an era with any kind of organized behavior that we could call an observer.

Before proceeding further, I should elaborate on what my ``philosophical" stance is towards anthropic arguments.  As the inventor of one of the first models, which was designed to explain the value of the c.c. on anthropic grounds\cite{davieslindebanks}, I am certainly not someone who rejects the scientific validity of such arguments outright.  However, I do believe in Albrecht's razor: ``The physicist who has the smallest number of anthropic arguments in her/his model of the world, wins"\cite{albrecht} .  More importantly, I believe that it's clear that many of the values of parameters in the standard model {\it cannot} be explained anthropically\cite{bdg}, especially if one allows for the existence of scales between the Planck scale and the scale of electroweak interactions. In particular, anthropic arguments cannot explain the existence of multiple generations of quarks and leptons and the bizarre pattern of couplings that determine their properties.  

Some authors attempt to get around these phenomenological problems by combining anthropic arguments with traditional symmetry arguments, but it is not at all clear that this makes sense.  In particular, in the most popular model for a distribution of parameters in the standard model, The String Landscape, the rules seem to imply that models with extra symmetries beyond the standard model gauge group are exponentially improbable\cite{dinesun}.   My own feeling is that, within the context of models where we insist on the standard model gauge group, the only way we could have a sensible anthropic explanation of what we see, would be if there were only one generation of quarks and leptons, and we relied on anthropic arguments to determine the weak scale\footnote{And this leads to unsuccessful predictions of the mass of the Higgs boson.} {\it and} insisted that a QCD axion with a GUT scale decay constant were the dark matter.  Apart from the problem of generations, this framework has depressing implications for high energy physics.  We will not find anything in accelerators we can imagine building.

Even this framework is not immune to criticism.  All known anthropic arguments, which rely on detailed properties of the particle physics we know, are about the properties of nuclear physics.   This is physics at the MeV scale, and we should really be formulating our arguments in terms of an effective field theory at that scale.  One might imagine arguing that nuclear physics would be irreversibly damaged if the underlying gauge theory did not consist of $SU(3) \times U(1)$ with the up and down quark masses and the QCD scale having values close to those in the real world.  However, one {\it cannot} imagine that the weak interactions affect nuclear and stellar physics in a way that cannot be mimicked by a host of four fermion interactions which are different than those in the standard model.  Thus, an honest anthropic argument, even one that makes the {\it a priori} assumption of our standard strong and electromagnetic gauge theory, cannot determine the form of the standard model Lagrangian.

Once one gives up the assumption of life based on the physics and chemistry we know, almost all bets are off.  We know too little about how physics determines biology, the possibility of radically different forms of organization and intelligence, or a host of other questions, to even pretend to make anthropic arguments in this wider context, except those which rely only on general properties of thermodynamics, and gravitation.

In HST, the questions like the nature of the low energy gauge group and the number of generations are determined by the fundamental commutation relations of the variables, and are not subject to anthropic selection among possible states in a given model.  Parameters like $n, \epsilon$ and $n_{BH}$, which characterize our cosmology may well be anthropically selected, subject to inequalities like $1 \ll n \ll N$ and $n_{BH} < n^{-3}, $ but in the models studied so far the cosmological horizon size is an input parameter.  In \cite{holocosm?} a more general model was proposed, in which $N$ is a variable.   That model is based on the existence of classical solutions of Einstein's equations in which a single horizon volume of de Sitter space is joined onto a stable black hole in the $p = \rho$ FRW model.  The black hole and cosmological horizons coincide.  
There is a completely explicit quantum mechanical HST model, whose coarse grained properties match those of this solution.  One can also construct more general solutions, in which such dS black holes, with varying horizon size, relative positions and velocities, move in the $p = \rho$ background.  This allows for anthropic selection of all of the parameters $N,n, \epsilon, n_{BH}$.  The nature of the anthropic arguments is quite different than conventional ones, because the HST formalism suggests a relation between $N$ and the scale of supersymmetry breaking.  I don't have space to discuss this in detail here, but the arguments give a plausible explanation of the data.

Another ``problem" that this generalized model solves is the existence of an infinite number of late time observers, ``Boltzmann Brains", whose experience is very different than our own.  One can arrange the initial positions and velocities of the black holes with different interior c.c., such that collisions always occur on time scales much longer than the current age of the universe, but much shorter than the time scale for production of Boltzmann Brains.  I don't consider this a major triumph, for the same reason that I don't think the BB problem is a serious one.  The difference between those two time scales is so huge that one can invent an infinite number of changes to the theory, which will eliminate the BBs without changing anything that we will, in principle, be able to measure.  BBs are a problem only to a theory which posits that the explanation for the low entropy initial conditions of the universe was a spontaneously fluctuation in a finite system.  In HST there is an entirely different resolution of the Boltzmann-Penrose question, so BBs are a silly distraction, which can be disposed of in a way that will never be testable.   Indeed, much of the structure of the HST model that allows for anthropic selection will remain forever beyond our reach, since it depends on initial conditions whose consequences are not causally accessible to us until our dS black hole collides with another.  Our universe then undergoes a catastrophe, on a time-scale of order $10^{61} L_P \sim 10^{10}$ years, somewhat analogous to what happens when Coleman-deLuccia bubbles collide.  After that time, the low energy effective field theory has changed and we will not be around to see a subsequent collision, if one occurs.  

As far as I can tell, any model which implements the anthropic principle will suffer from similar problems.  It must predict many things, which no local observer can ever observe.  This is the reason that I subscribe to Albrecht's razor: a non-anthropic explanation for a fact about the universe can be tested more thoroughly than an anthropic model can.  This doesn't mean that one can ignore the possibility that some of what we observe depends on accidental properties of particular meta-stable states of a system larger than anything we can observe.  I have spent years trying to find a more satisfactory explanation of the value of the c.c., and I conclude that this will not be possible.  The HST model also suggests that $n, n_{BH} $ and the precise form of the early FRW metric during the transition from inflation to the dilute black hole gas phase, are also free parameters,which characterize the initial state.  They are subject to a combination of entropic and anthropic pressures, and I believe they can be pretty well pinned down by these arguments.

On the other hand, in HST the low energy particle content of the model is determined by the super-algebra of the variables and is fixed once and for all.
Different spectra correspond to different candidate models.  My hope is that very few of these candidate models actually lead to mathematical consistency.
We're familiar from string theory and low energy effective field theory that most candidate models of quantum gravity are inconsistent.  Even string models with $4$ dimensional $N = 1$ supersymmetry, many of which are consistent to all orders in perturbation theory, are expected to be actually consistent for at most discrete values of the various continuous parameters that label these models in perturbation theory.  Indeed, generically, there is no reason to believe that the perturbation series is accurate at any of the consistent points. The string perturbation series determines quantities, the scattering amplitudes in Minkowski space, which simply do not exist unless the point in parameter space at which the model exists preserves both supersymmetry and a discrete phase rotation (R) symmetry, which acts on the supersymmetry generators.  Such points are very sparse in the space of all parameters.

The theory of SUSY breaking in HST\cite{susyholo} imply that in the limit that the c.c. is taken to zero, the model becomes super-Poincare invariant, has a discrete R symmetry\footnote{A discrete R symmetry is a discrete symmetry group which acts on the fermionic generators of the SUSY algebra.}, and no continuous moduli.  There can be no examples of such a model in perturbative string theory, since the string coupling itself appears to be a continuous parameter.  General arguments in effective super-gravity imply that such models are rare, corresponding to solving $p + 1$ equations for $p$ unknowns.  Furthermore, in perturbative string theory, one can look at the analogous problem of fixing all parameters besides the string coupling and finding a model with a discrete R symmetry.  Again, such models are rare.   It is thus plausible to guess that consistent HST models with vanishing c.c. are rare, and the gauge groups and matter content of these models, as well as their parameters, highly constrained.  It is not out of the question that we'll be able to find arguments that the standard model is the unique low energy gauge theory, which can arise at such a point.

\section{Conclusions}

On an intuitive level, HST models of cosmology are quite simple.  The basic principle behind them is that bulk localized excitations in a finite area causal diamond are constrained states of variables living on the boundaries of the diamond.  The low entropy of the state of the early universe is explained by the necessity of having such localized excitations in an observable universe: the {\it topik\`{e}sthropic principle}.  The initial state with maximal {\it localized} entropy 
is a collection of black holes.  A competition then ensues between black hole coalescence and black hole decay, which must end in most of the black holes decaying if the universe is every to develop complex structures.  This requires a fairly uniform dilute gas of black holes.  Absolute uniformity is impossible, because the black holes are finite quantum systems undergoing fast scrambling dynamics.  This leads to fluctuations in the mass and angular momenta of the black holes, of order \begin{equation} \frac{\delta M}{M}\sim \frac{\delta L}{L} \sim S^{- 1/2}, \end{equation} where $S$ is the black hole entropy.  These are the fluctuations we see in the sky.

What is remarkable is that this model, actually has all the features traditionally associated with inflation.  The two crucial ingredients in this conflation of apparently different models, are the different time slicings associated with the HST model, and FRW space-time, and the necessity, in the HST model, to describe the evolution of the black holes, before they enter the horizon, as decoupled quantum systems, of fixed entropy.  Each decoupled system is the same as the HST model of dS space (with the inflationary Hubble radius, $nL_P$), so, if we want a homogeneous distribution of black holes, then FRW time slices up to conformal time $\eta_0 / 2$ must be described as a collection of decoupled quantum systems, of fixed entropy.  $\eta_0$ is the point in conformal time to which our universe converges as the proper time along trajectories goes to infinity.  This system thus corresponds to the conventional picture of an inflationary universe as many independent horizon volumes of dS space.
The number of e-folds is fixed, with the precise value of $N_e$ dependent on $n_{BH}$, the primordial density of black holes at the end of inflation.  That number density also determines the reheat temperature of the universe after black hole decay.  It is bounded from above by $n_{BH} < n^{-3}$ .  The size of primordial scalar fluctuations is $ Q \sim (n\epsilon)^{-1} $, where $\epsilon = \frac{\dot{H}}{H^2} $ is a slow roll parameter.  For $\epsilon \sim 0.1$ the CMB data tells us that $n \sim 10^6$ and this implies that the reheat temperature is less than $10^7 - 10^8$ GeV. 

I also sketched the argument that the HST curvature fluctuations should be approximately $SO(1,4)$ invariant, which is enough to account for the data, with detailed features of the scalar spectrum fit to $H(t)$ the background FRW metric at the end of inflation.  This is just as in conventional inflation models, but those models make much more specific assumptions about the state of the quantum system under discussion, assumptions which amount to a massive fine tuning of initial conditions.  They also have to go through an elaborate discussion, rarely touched on in the mainstream inflation literature, to justify why quantum fluctuations in a quantum ground state decohere into a probability distribution for the classical curvature fluctuations.

In contrast the HST model has fluctuations arising from localized quantum systems with a huge number of states, with the curvature interpreted as in \cite{ted} as a hydrodynamic average property.  In more familiar terms, the fluctuations are fluctuations in mass and angular momentum of mesoscopic black holes\footnote{Black holes for which quantum fluctuations are not entirely negligible, though already decoherent. $1/n$ is not negligible, but $e^{- n^2} $ is.}.  This model also has fine tuning of initial conditions, but it is the minimal tuning necessary to obtain a universe with localized excitations which are not black holes.

Unfortunately, current cosmological data do not allow us to distinguish between these two very different models or many other more exotic field theory models with multiple fields, strange forms of kinetic energy, {\it etc.}.  The most likely observational distinctions to be measured in the near future will be short wavelength gravitational waves.  In the distant future, measurement of the tensor three point function might definitively rule out quantum field theory as the source of CMB fluctuations.

\end{document}